**Technomolecular Materials: 3D-Printed 2D-Nanosheets with Self-Patterned Electrodes**

Hicham Hamoudi*, Sara Iyad Ahmad, Atef Zekri, Kamal Youcef-Toumi, Vladimir A. Esaulov

**Affiliations:**

H. Hamoudi*, S. I. Ahmad, A. Zekri: Qatar Environment and Energy Research Institute, Hamad Bin Khalifa University, 34110, Doha, Qatar.

*Corresponding author Email: hhamoudi@hbku.edu.qa, hichamhamoudia@gmail.com

K. Youcef-Toumi: Mechatronics Research Laboratory, Massachusetts Institute of Technology, MA 02139, USA.

Vladimir A. Esaulov: Institut des Sciences Moléculaires d'Orsay, UMR 8214 CNRS-Université, bât 520, Université Paris Sud, Université Paris Saclay, Orsay 91405, France.

**Abstract:**
Building on our prior work, where our team transcended self-assembled molecular monolayers (SAMs) research from a 2D configuration to 3D-structured materials and successfully introduced the molecular self-assembled 3D printer to fabricate technomolecular materials—hybrid carbon/metal nanosheets that mimic biological self-assembly through cooperative organic-inorganic interactions—these materials promise advances in nanotechnology by enabling seamless integration of molecular systems with metallic electrodes. Here we show that electron-beam irradiation induces direct self-patterning of silver fractal nanoelectrodes on the technomolecular nanosheets, with formation influenced by molecular structure: saturated variants yield localized nanoparticles, while conjugated ones produce propagated fractals via electron delocalization and cross-linking. In-situ transmission electron microscopy reveals dynamic diffusion aggregation mechanisms, allowing controlled circuit patterns through resist-free electron-beam lithography. This approach advances flexible electronics, bioelectronics, and energy conversion, including fractal antennas and unclonable identifiers.

**Keywords:** technomolecular; self-assembly; fractal; nanosheet; in-situ TEM; 3D printing

## 1. Introduction

The human brain, an intricate and highly complex organ, has long inspired researchers in their pursuit of advanced nano electronic systems. A key aspect of the brain's complexity lies in its 3D self-assembly process, which underpins the spontaneous generation of its biological matter and its ractal neural networks [1]. This aspect frames the brain as a sophisticated organic system intertwined with technological aspect, epitomizing 'technorganic material' a mix of technological and biological elements. Fractal architectures, inherent in the neural system, offer numerous advantages such as effective surface area utilization and signal transfer, making them an attractive design paradigm for a variety of technological applications [1-2]. In fact, fractal-inspired designs have driven innovations in processors, antennas, solar-cell structures, and energy conversion devices [3-9]. Recent studies have explored the development of fractal-like electrical networks to improve implant/neuron communications [7].

The central aim of Nanoarchitectonics is to construct functional material systems by designing molecules through the self-assembly of organic molecular units mimicking biological systems. Nonetheless, the integration of these sub-nanometer molecular assemblies with traditional electronic systems, particularly incorporating metallic electrodes, continues to present a substantial challenge[10-12]. We introduce Technomolecular materials as means to overcome challenges in molecular electronics and usher in the next generation of nanotechnologies [11]. This class of materials draws inspiration from properties and structures found in living organisms, harnessing cooperative interactions between organic and inorganic elements, and resulting in materials with distinctive and desirable traits. Technomolecular materials consist of hybrid carbon/metal nanosheets that are synthesized by a molecular 3D printing technique reported recently by Hamoudi [13]. During this 3D printing technique, the organic carbon molecules are self-assembled into monolayers joined together by metallic atoms that are introduced during the printing process, eventually creating a nanosheet that consists of a periodic and continuous carbon-metal-carbon multilayer structure endowed with exceptional self-healing properties [13].

In this study, we unveil another aspect of these materials emerging upon exposure to an electron beam, leading to localized modifications and resulting in the self-assembled fractal patterning of metallic nanostructured electrodes enabling resist-free lithography. Our investigation focuses on understanding how molecular electronic properties influence the formation of these metallic fractal nanostructures using *in-situ* Transmission Electron Microscopy (TEM) experiments. Thus, a high-energy electron beam (e-beam) was used to locally modify two different conjugated and saturated nanosheets and induce patterning of metallic nanostructures at the same time. Furthermore, we address a critical challenge related to the nonlinear propagation of fractal electrodes by designing nano-trap pathways to regulate the formation of these nanostructures. Finally, we demonstrated the feasibility of resist-free electron-beam lithography for fabricating complex circuit patterns directly on the printed nanosheets. The reported work is novel and provides a forward pathway to engineer the next generation of molecular and flexible electronics, bioelectronics, and energy conversion including the next generation antenna technologies. This strategy for designing top-electrodes is expected to play a pivotal role in integrating molecular electronics into the heart of computers.

## 2. Results & Discussion

### 2.1 Morphology of the as-Printed Saturated C9/Ag and Conjugated BPD/Ag Nanosheets

Hamoudi *et al*. [13]. investigated the process of 3D printing by means of self-assembly of individual molecules [14]. Details on the self-assembled molecular 3D-printing synthesis of 2D nanosheets can be found in [13]. Briefly, the concept behind the printed nanosheets is to use dithiol molecules as building blocks and metal ions as mediators connecting the thiol-based monolayers together. First, a high-quality dithiol self-assembled monolayer (SAM) on a metallic substrate was prepared and immersed into a $AgNO_3$ water solution [13]. Because of the affinity of the thiol molecule towards metals, the surface of the SAM on the substrate grafted and bonded with the $Ag^+$ atoms in water. The dithiol molecules dispersed in hexane were continuously injected through a contact angle nozzle while simultaneously pulling the nozzle upwards. The immiscibility between the hexane and water prevented the thiol molecules from dispersing into the water but allowed them to bond with the metal atoms at the water-hexane interface. This resulted in the formation of a hexane bubble that separated both liquids and was encapsulated with the hybrid dithiol/Ag nanosheet. On the other hand, the continuous movement of the nozzle directed the ordering of the formed monolayers resulting in a periodic and continuous multilayer self-assembled structure, see schematic graph in [13]. As a result, hybrid metal/organic nanosheets are formed, see Fig. 1 and Movie V1 in the Supporting Information (SI) for the printing of the saturated 1,9-nonane-dithiol (C9)/Ag nanosheet.

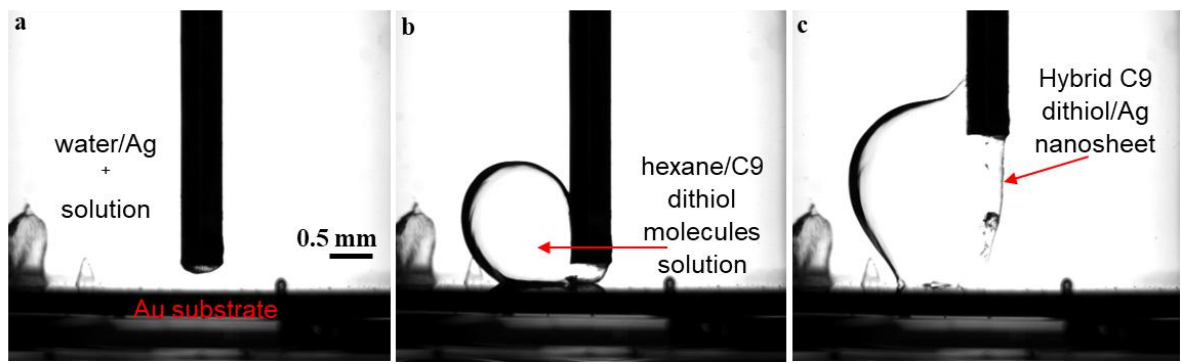

Figure 1. A sequence of optical pictures taken during the printing process of the C9/Ag nanosheet. Starting from (a) to (c), the nozzle was brought into proximity with the SAM decorated gold substrate in a water-Ag$^+$ solution. Then, the C9 molecules dispersed in hexane were continuously injected through the nozzle while simultaneously pulling the nozzle upwards. The C9/Ag nanosheet was formed at the water-hexane interface.

The defect-free, nanometric, flexible, wrinkled, and folded sheet-like morphology of the as-printed C9/Ag nanosheet is shown in the bright-field TEM image, see Fig. S1 in SI. A detailed peak-force atomic force microscopy analysis reported in [13] showed that the thickness of the as-printed nanosheet is ~5nm. Furthermore, TEM elemental mapping of the nanosheet, see Fig. S1b-S1e in SI, indicates the homogenous distribution of the sheet's constituents including Ag, C, and S, which is a characteristic feature of the self-assembly mechanism indicating the uniformity of the material.

FTIR spectroscopy can reveal specific spectroscopic fingerprints shedding light on the molecular organization and order of the as-printed nanosheets [15]. The FTIR spectra of the as-printed 5,5′-bis(mercaptomethyl)-2,2′-bipyridine BPD/Ag nanosheet as compared to its reference BPD powder can be seen in Fig. S2. The spectra of the BPD/Ag nanosheet display anti-symmetric $\upsilon_{as}(CH_2)$ and symmetric $\upsilon_s(CH_2)$ stretch peaks at 2919 and 2850cm$^{-1}$, respectively. These values are the fingerprint of closely packed alkyl chains that display crystalline ordering [16]. This is further confirmed by the red-shifted stretching frequency $\upsilon_{as}(CH_2)$ of the reference BPD powder at 2929cm$^{-1}$ which is typical of SAMs with lower crystallinity in their chain packing. These findings agree with those of ordered SAMs reported in the literature and indicate that the printed nanosheet displays ordered carbon chain characteristics [15,17], consistent with the self-assembly mechanism. Furthermore, the FTIR spectra of the reference BPD powder exhibits a strong peak at 2558cm$^{-1}$ representative of S-H vibrations from the thiol groups [18], which is absent from the spectra of the printed BPD/Ag nanosheet. This corroborates the building block mechanism where the S-H in each BPD molecule was substituted with the S-Ag bond connecting the self-assembled BPD monolayers together and forming the backbone of the nanosheet during the 3D-printing process.

*2.2 Metallic Nanostructure Formation Under Electron-Beam Irradiation*

*2.2.1 Behavior of Saturated C9/Ag Nanosheet Under Electron-Beam Irradiation*

First, we investigated the capability of the e-beam to induce metallic crystallization of the $Ag^+$ in the saturated C9/Ag Technomolecular nanosheet. The change in the nanosheet's morphology and structure before and after e-beam irradiation is shown in Fig. S3a and S3b, respectively. Initially, a lower electron dose of $800A/m^2$ was used to irradiate the red-circled area, see Fig. S3b. After 20 seconds of irradiation, small dark features emerged. The irradiated area was then magnified, Fig. S3c, and further irradiated for 2 minutes. The resulting nano-features exhibited a homogeneous distribution across the irradiated region with an average size of 10nm. The Selected Area Electron Diffraction (SAED) pattern shown in Fig. S3d was indexed as FCC Ag, verifying the metallic nature of these Ag nanostructures, without any secondary diffraction rings observed. STEM-HAADF elemental mapping shown in Fig. S4 confirmed the formation of Ag nanoparticles (NPs) and suggested that the Ag NPs were encapsulated with C in the nanosheet.

The rapid formation of Ag NPs indicates the high sensitivity of the nanosheet to the e-beam and suggests the transformation of the $Ag^+$ connecting the C9 monolayers to homogeneously distributed Ag NPs in the irradiated areas of the nanosheet. While there has been extensive research on synthesizing nanostructures through e-beam irradiation [19-21], a comprehensive understanding of the underlying interactions remains speculative [22]. Generally, it is known that e-beam/matter interactions result in electronic excitation and ionization, which can lead to bond breakage, atomic displacement, sample charging, or sample heating [20,23-24].

An *in-situ* TEM investigation was conducted to gain further insight into the nucleation and growth mechanisms of the Ag NPs on the C9/Ag nanosheet. A stationary e-beam was focused on one area in the C9/Ag nanosheet and a sequence of TEM images with different magnifications were taken throughout the analysis, see Fig. S5, and V2-V4 in SI. It can be observed from Fig. S5a-S5b that the NPs exhibited burst nucleation best described by the LaMer mechanism [25]. High-energy e-beam irradiation caused bond cleavage within the organic-based nanosheet, leading to the release of $Ag^+$ that were reduced to Ag and aggregated homogeneously in the irradiated area. The Ag nuclei gradually grew by capturing other silver atoms, turning into clusters, and eventually into NPs, see V2-V4 in SI. Growth by coalescence occurred when two NPs were in proximity, see yellow circles in Fig. S5c-S5d, and V3 in SI, with rapid necking in a fluid-like motion occurring before merge completion. Surface diffusion was identified as the main mechanism of metallic NPs growth by coalescence [26] facilitated by the energy release through the reduction in the necking surface area. Ag NPs mainly nucleated and grew confined in the areas irradiated by the e-beam, see Fig. S5h-S5i.

With continued growth, a tear emerged near the upper-left side of the larger particle and created a hole, as indicated by the white arrows, see Fig. S5f and V4 in SI. Reducing the magnification indicated that the rest of the sheet remained intact while the initial tear grew with prolonged irradiation to reach the adjacent NP, see Fig. S5g. Organic nanosheets, particularly those composed of aliphatic/saturated molecules such as alkane thiols, tend to be unstable when exposed to e-beam irradiation. The intensive electron interactions result in the cleavage of the C–H and C-C bonds inducing desorption of the material [23-24,27-28].

Until the consumption of $Ag^+$ ions and the growth saturation of Ag NPs in the irradiated region, the nanosheet remained intact, suggesting that part of the energy from the initial e-beam irradiation was utilized in the ionization, diffusion, and growth of Ag NPs. After which, the accumulation of static electric charges in the organic nanosheet within the exposed area prompted charging and subsequent damage to the nanosheet. It can be concluded that the formation of the metallic Ag NPs contributed to the stability of the saturated nanosheet against irradiation. Similar results were reported by Khan *et al*. [29] as they investigated the effect of Au NPs on the stability of MWCNT under e-beam irradiation.

In addition to $Ag^+$ diffusion and NPs coalescence, a third growth mechanism of Ostwald ripening was observed, see Figs. 2a-2c and V5 in SI. Under continuous electron irradiation, the two larger NPs (1 and 2) continuously approached each other, fused, and merged across the interface with irradiation time. On the contrary, the smaller NPs (3 and 4) reduced in size with irradiation time until they disappeared, see white arrows in Fig 2b. The occurrence of several growth mechanisms have been observed during *in-situ* TEM growth of metallic NPs [20,30].

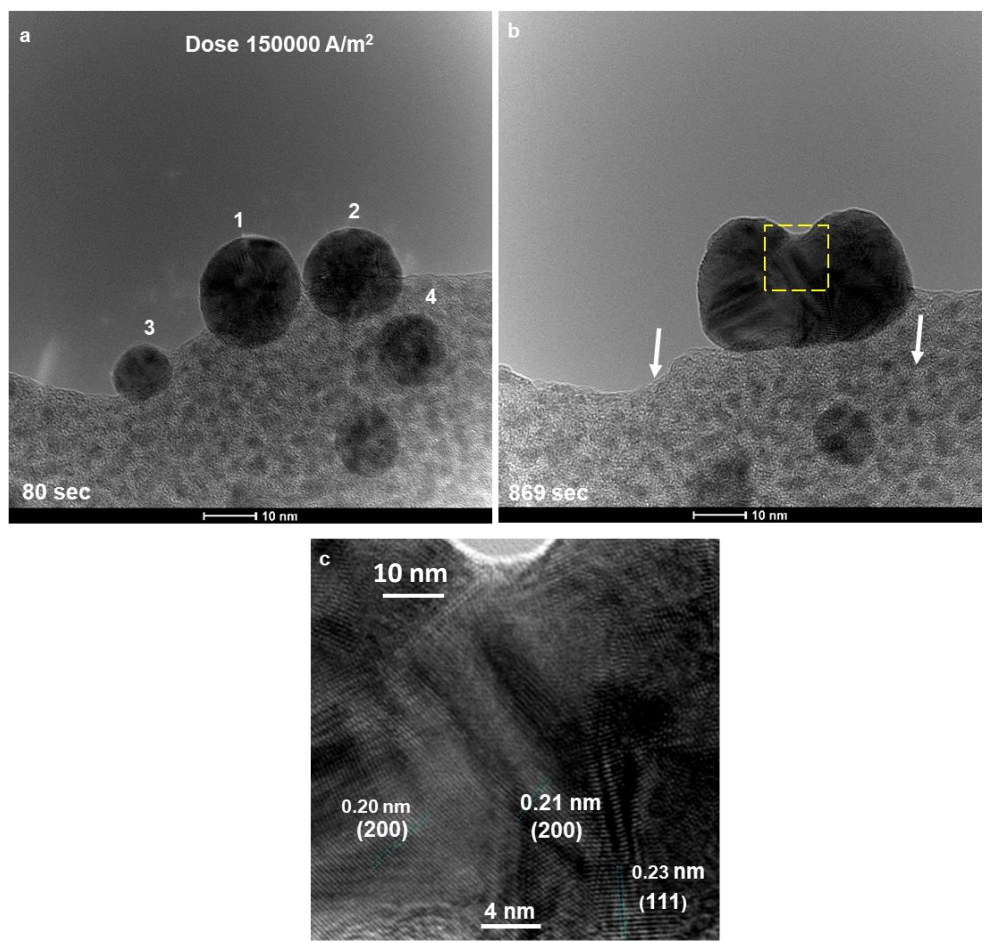

Figure 2. TEM images of Ag NPs grown by e-beam irradiation and located at the edge of the hybrid C9/Ag nanosheet. The white arrows in (b) point to the initial locations of particles 3 and 4 in (a) before they vanished. (c) Magnification of yellow square in b.

*2.2.2 Behavior of conjugated BPD/Ag Nanosheet Under Electron-Beam Irradiation*

When exposed to e-beam irradiation, the response of the conjugated BPD/Ag Technomolecular nanosheet was distinct from its saturated counterpart. At a low electron dose of 200A/m², the NPs formed and grew in a ramified chain-like reaction resulting in the formation of a fractal network over time, see Fig. 3. The irradiation-induced metallic nanofractals consisted of long central metallic backbones and sharp secondary and ternary branches with a fractal width of ~10nm. The fractal-dimension of the fractal pattern shown in Fig. 3a is ~1.6 as determined by the box-counting method using the Fiji-ImageJ software [31], see Fig. S6, which is consistent with the 1.7 value of a typical diffusion limited aggregation (DLA) fractal pattern [32]. It is interesting to note that the fractal dimension of retinal neurons is ~1.7 [33].

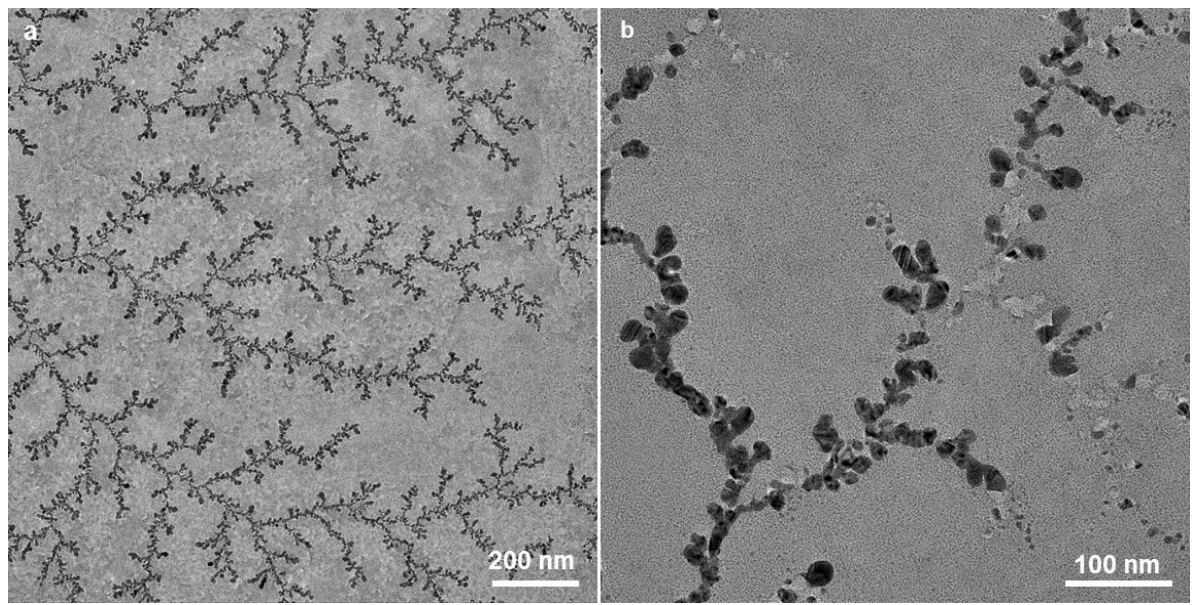

Figure 3. BF-TEM images of the printed BPD/Ag nanosheet taken immediately after exposure to a low-dose e-beam. (a) fractal formation of Ag nanoelectrodes, and (b) magnification of the fractal nanostructure.

With increased magnification and prolonged irradiation, no tearing or damage was seen in the conjugated nanosheet, see Fig. 3b. Instead, the higher magnification and its associated stronger electron bombardment resulted in the dissolution of the NPs as they disappeared and restored at different places repeatedly, leaving behind a trace of a cavity in the nanosheet. Under the same synthesis procedure and similar irradiation conditions, the only difference in both nanosheets was the electronic structure: saturated or conjugated. Hamoudi *et al*. [34] studied the effect of electron irradiation on the electronic transport properties of saturated and conjugated molecular SAMs. The XPS results revealed that saturated alkane-thiol SAMs suffered significant damage by electron bombardment, whereas conjugated SAMs remained structurally intact through the cross-linking of their aromatic molecules. This observation explains why the aromatic BPD nanosheet remained undamaged during prolonged irradiation, as the cross-linking of its aromatic rings helped overcome the cleavage of molecular bonds.

The conductivity across the as-printed C9/Ag and BPD/Ag nanosheets was measured for further investigations. The current-voltage transport measurements of both nanosheets were conducted using scanning probe microscopy (SPM), see Fig. S7. The largest current density of 286A/cm$^2$ was obtained for the BPD/Ag nanosheet, which is more than 1000 times larger than that of its aliphatic counterpart. Rooted in its poor conductivity and minimum charge mobility, the C9/Ag nanosheet suffered significant charging (indicated by the presence of a minimum current at 0 bias) [35]. This charge buildup supports the observed sheet damage seen in Fig. S5f. Once the Ag$^+$ were consumed and localized Ag NPs formed, the high density of localized static electrons triggered the nanosheet's damage.

In an opposite scenario, the lifetime of charge buildup persisted for shorter times in the BPD nanosheet as indicated by its conductivity results. This is due to the electron delocalization characteristics of the BPD electrons, which enabled the accommodation of excess electron-irradiation induced charging through charge propagation. The experimentally observed stability of the aromatic nanosheet under electron-irradiation is therefore attributed to charge propagation, cross-linking of its aromatic chains, and formation of metallic fractals.

*2.3 Direct Assembly and Control of Metallic Electrodes on the Self-Assembled Carbon-Based Nanosheets*

The feasibility of generating metallic nanopatterns by direct e-beam irradiation in both saturated and conjugated nanosheets was investigated in this section. Since the metallic NPs in both nanosheets form spontaneously in the irradiated areas, as observed earlier, creating different patterns in the nanosheets must be done by defining the areas in which the NPs do not form. To do so, patterning is done in two steps: exposing the nanosheets to a high-dose irradiation to expel the Ag$^+$ from the irradiated areas and define the required patterns, followed by exposing the nanosheets to a low-dose irradiation to induce the gradual and homogenous growth of metallic nanostructures in the remaining non-patterned areas. The different behaviors of both nanosheets upon patterning can be seen in Figs. 4 and 5.

Several patterns were created in the C9/Ag nanosheet, see Fig. 4(a-c). A stationary e-beam with a dose of ~100,000A/m$^2$ was used in Fig. 4a, while in Fig. 4b, a moving beam was used to create a line pattern. The moving e-beam with a spot size of ~100nm and a dose of ~80,000A/m$^2$ was moved vertically across the nanosheet at a rate of 40nm/s. In Fig. 4a, the Ag NPs formed a ring pattern at the periphery of the irradiated area, while in Fig. 4b, metallic NPs formed parallel to both sides of the e-beam, creating a line profile in the nanosheet. A more complex trap-like pattern was created in Fig. 4c using both stationary and moving beams. Details on the creation of this pattern are summarized in Methods. Similarly, localized NPs formed at the borders of the irradiated areas, see Fig. 4c.

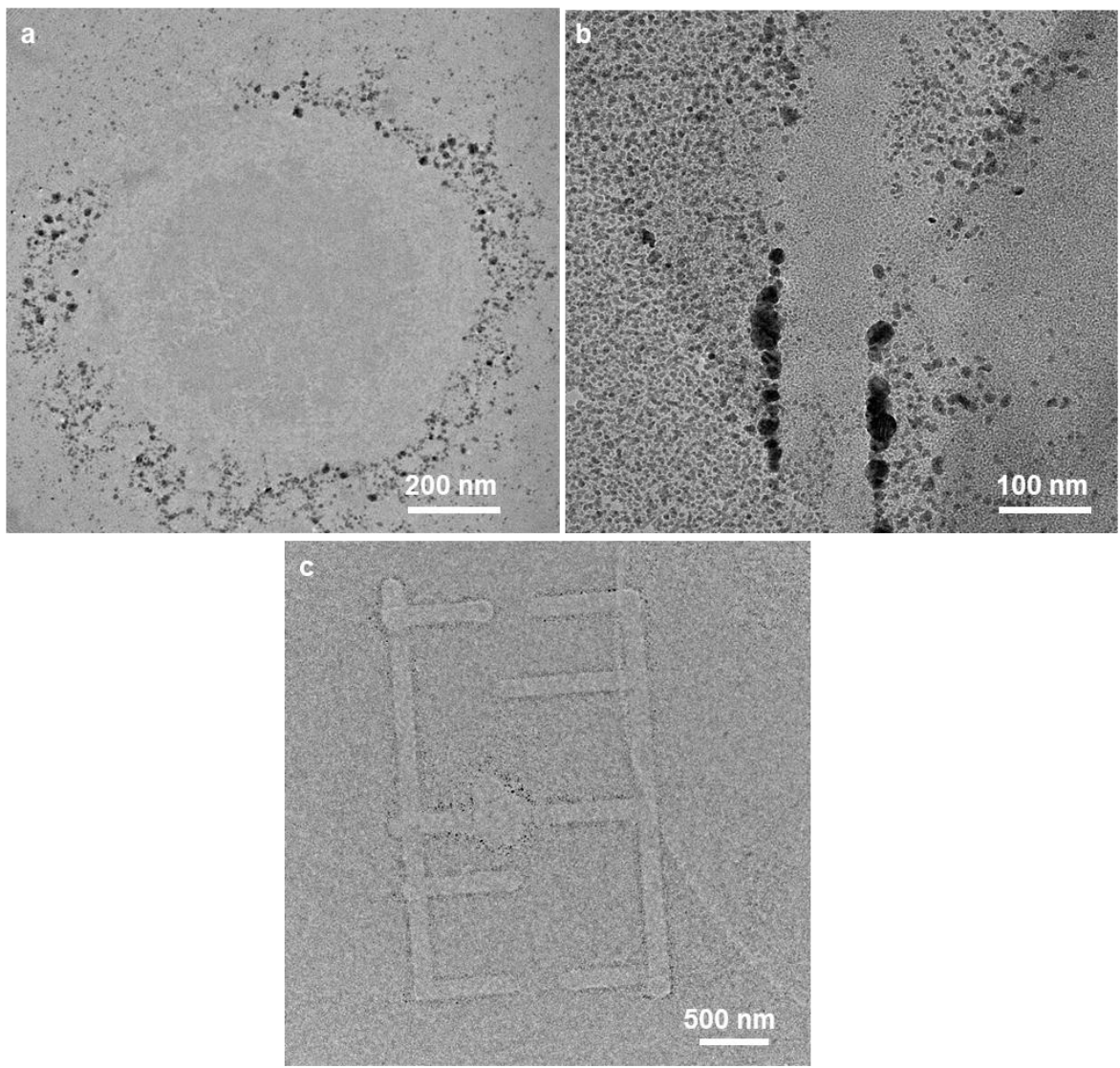

Figure 4. Creating different patterns in the saturated C9/Ag nanosheet using e-beam. (a-c) TEM images show the ability to create (a) circular, (b) linear, and (c) complex patterns onto the C9/Ag nanosheet.

Exposing the C9/Ag nanosheet to a sudden high-dose stationary e-beam ~80,000-150,000A/m$^2$ resulted in the sudden release of Ag$^+$ ions by radiolytic interactions and induced the loss of negative charge through the emission of secondary and auger electrons in the irradiated area [36-37]. This created a positive potential and a built-in electric field that expelled the positive Ag ions from the illuminated area to neutralize the local charge imbalance. At the same time, negative charges, be it escaped electrons or negative radicals in the nanosheet, were attracted to the expelled Ag$^+$ meeting at the realm of the illuminated area. This is confirmed by the presence of the Ag NPs at the periphery of the irradiated area away from the beam center indicating the migration of Ag$^+$ prior to the nucleation of Ag NPs. This phenomenon of NPs formation at the realm of the irradiated e-beam has been reported for Au NPs in liquid cells [37] and Ag, Cu, and Na in glass substrates [38-40], and is referred to as damage-indued electric field (DIEF) [40-41].

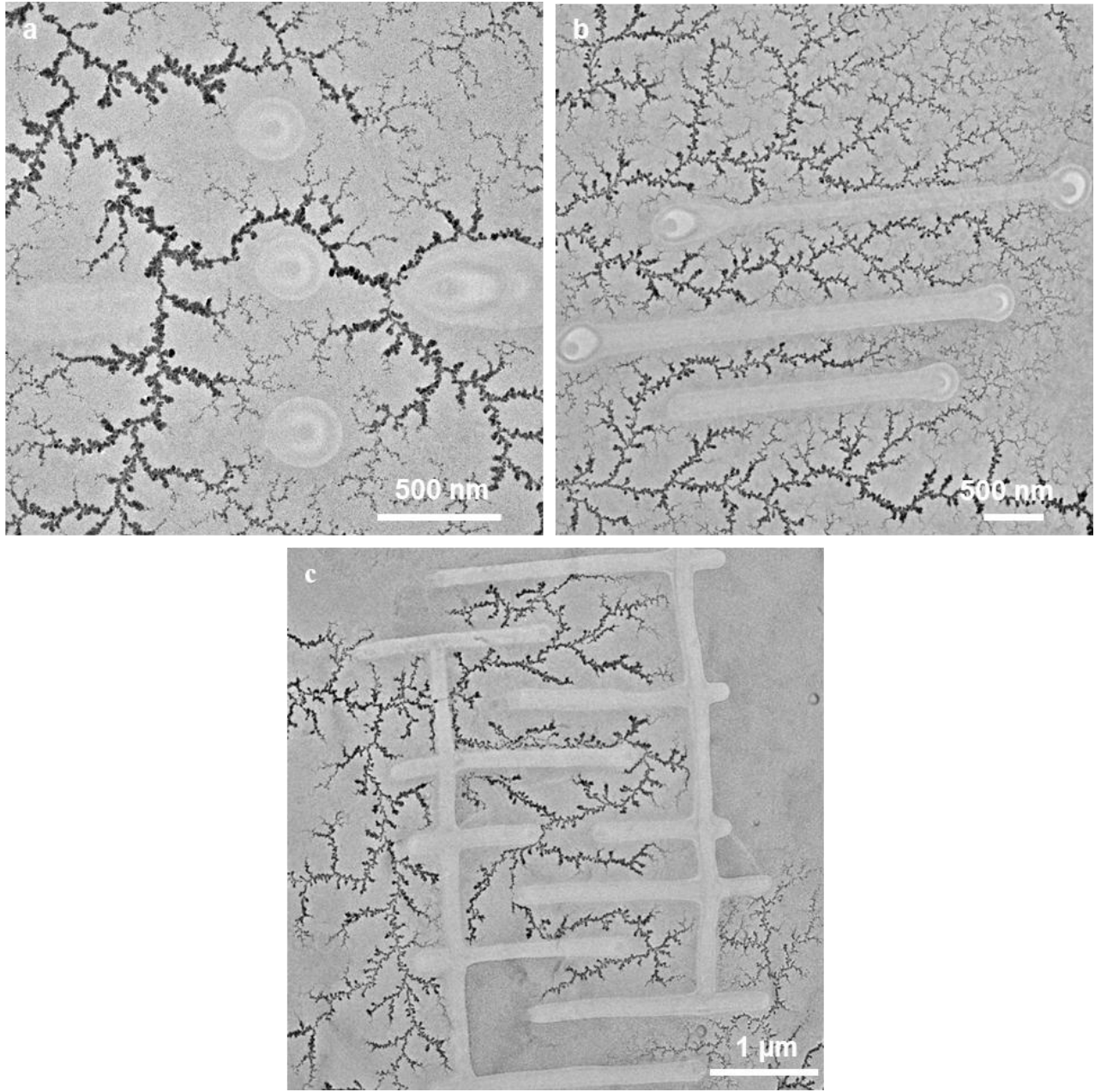

Figure 5. Creating different patterns in the conjugated BPD/Ag nanosheet using e-beam irradiation and controlling the paths of the fractal nanostructures in the nanosheet. (a-c) TEM images show the ability to control the growth path of the metallic Ag nanofractals into (a) curved, (b) linear, and (c) complex patterns.

Due to their nonlinear nature, fractals are notoriously difficult to control [42-43]. To regulate the behavior of a fractal electrode, trap paths were created to guide its movement, see Fig. 5(a-c). After creating the desired pattern, the magnification and electron dose were reduced, and metallic fractals formed immediately. Depending on the created patterns, the growth of the fractal nanostructures was controlled to produce linear, circular, or complex patterns, as shown in Fig. 5(a-c) and the *in-situ* recordings in V6-V8 in SI. The propagation of the Ag fractals mostly avoided the irradiated regions and occurred around the initially patterned path. This can be attributed to the cross-linking induced by irradiation inside the irradiated regions [34] during the initial high-dose patterning, which created a lack of free-electrons preventing the formation of fractals. Furthermore, the nucleation of Ag NPs in the BPD/Ag nanosheet was not localized at the borders of the patterned paths. Instead, the fractals propagated from outside of the irradiated area until they reached the created trap pattern.

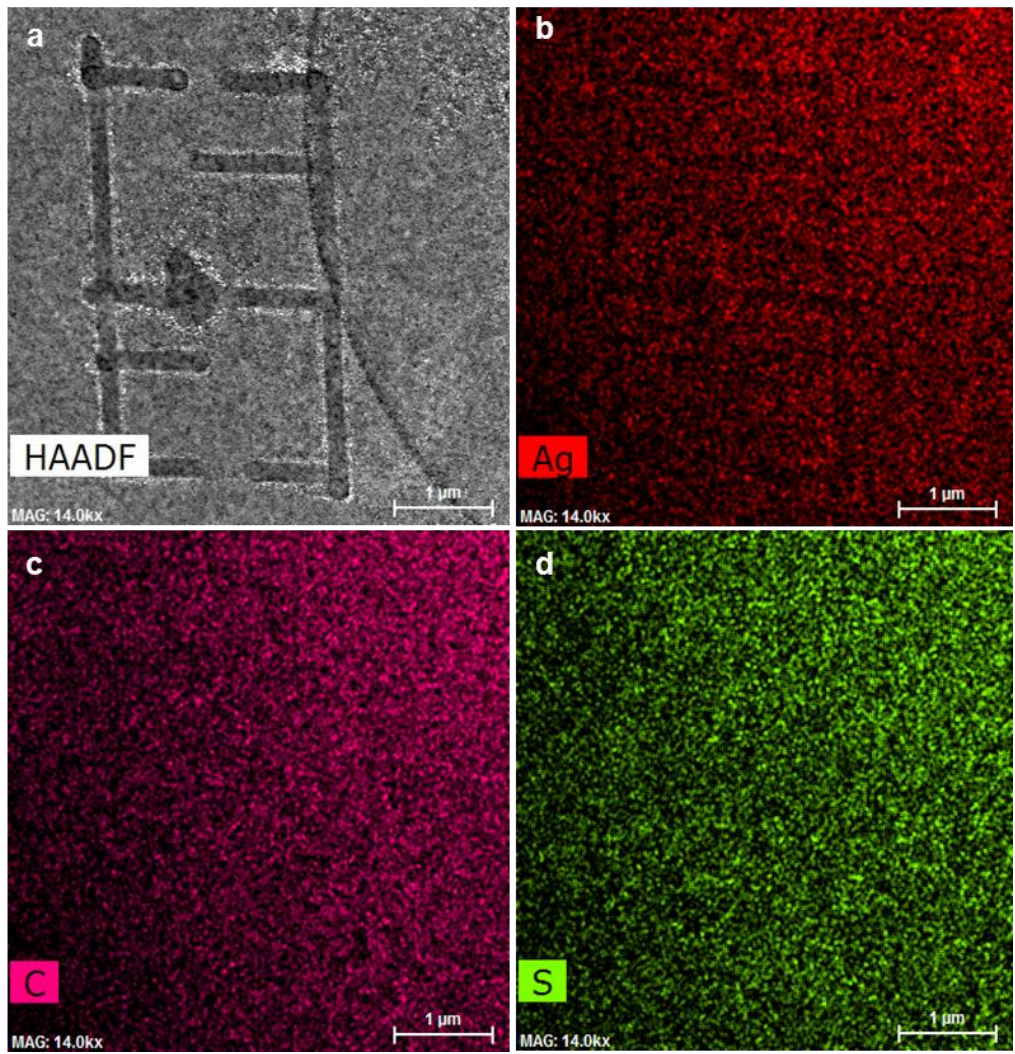

Figure 6. TEM elemental mapping of the saturated C9/Ag nanosheets. HAADF-TEM of complex trap-like patterns created in the (a) C9-dithiol/Ag nanosheet using high electron-doses of ~100,000 A/m2, followed by low electron-

S12

dose irradiation of ~200 A/m2 to initiate the Ag NPs formation. The corresponding elemental mapping of Ag in (b), C in (c), and S in (d).

Interestingly, the patterned areas on the BPD/Ag nanosheet appeared lighter in color compared to the non-patterned areas, whereas the C9/Ag nanosheet exhibited less color variance. The corresponding mappings for Figs. 4c and 5c can be found in Figs. 6 and 7, respectively. The concentration of Ag$^+$ was higher in the patterned areas of the C9/Ag nanosheet than in the BPD/Ag nanosheet. This is attributed to the expulsion of metal ions caused by the high-dose e-beam irradiation, resulting in a lower metal ion concentration in the irradiated regions compared to the surrounding areas. This provides further evidence that charges (either electrons or ions) are more mobile in the conjugated BPD nanosheet than in its saturated counterpart, which was further facilitated by the DIEF.

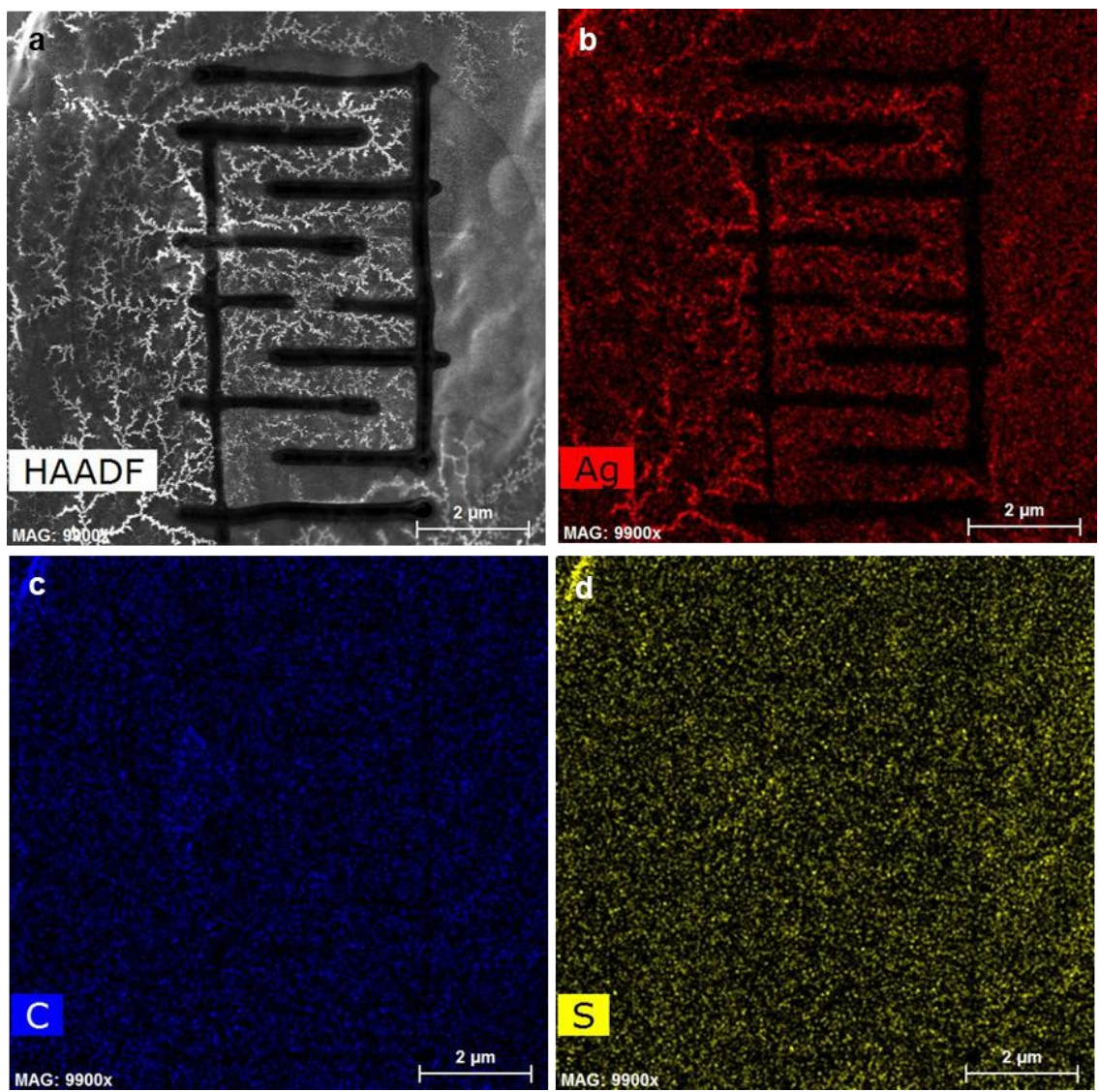

Figure 7. TEM elemental mapping of the conjugated BPD/Ag nanosheets. HAADF-TEM of complex trap-like patterns created in the (a) BPD /Ag nanosheet using high electron-doses of ~100,000 A/m$^2$, followed by low electron-dose irradiation of ~200 A/m$^2$ to initiate the Ag NPs formation. The corresponding elemental mapping of Ag in (b), C in (c), and S in (d).

Figure 8 is a lower magnification TEM image taken for the BPD nanosheet after several pattern creations, e-beam irradiation, and fractal formation. It is evident in Fig. 8 that the initiation of the fractal nanostructures stemmed from the edge of the surrounding TEM Cu grid. Deciphering the exact mechanism behind the formation of these fractals and providing a qualitative description of their origin is a daunting task. It is hypothesized that the fractal formation of the Ag nanofractals is related to the propagation of electrons within the nanosheet. We believe that the formation of metallic fractals in the BPD nanosheet resembles the Lichtenberg patterns that appear on human skin following a lightning strike. This is due to the electric discharge of air and the propagation of charges between two opposing potentials [44]. Feynman [45] explained that lightning is a complex phenomenon that originates from a negatively charged cloud over a positively charged ground. The cloud forms a negative "step leader" that branches outward as it seeks a connection to the ground, creating an ionized path in the air and resulting in the visible lightning bolt.

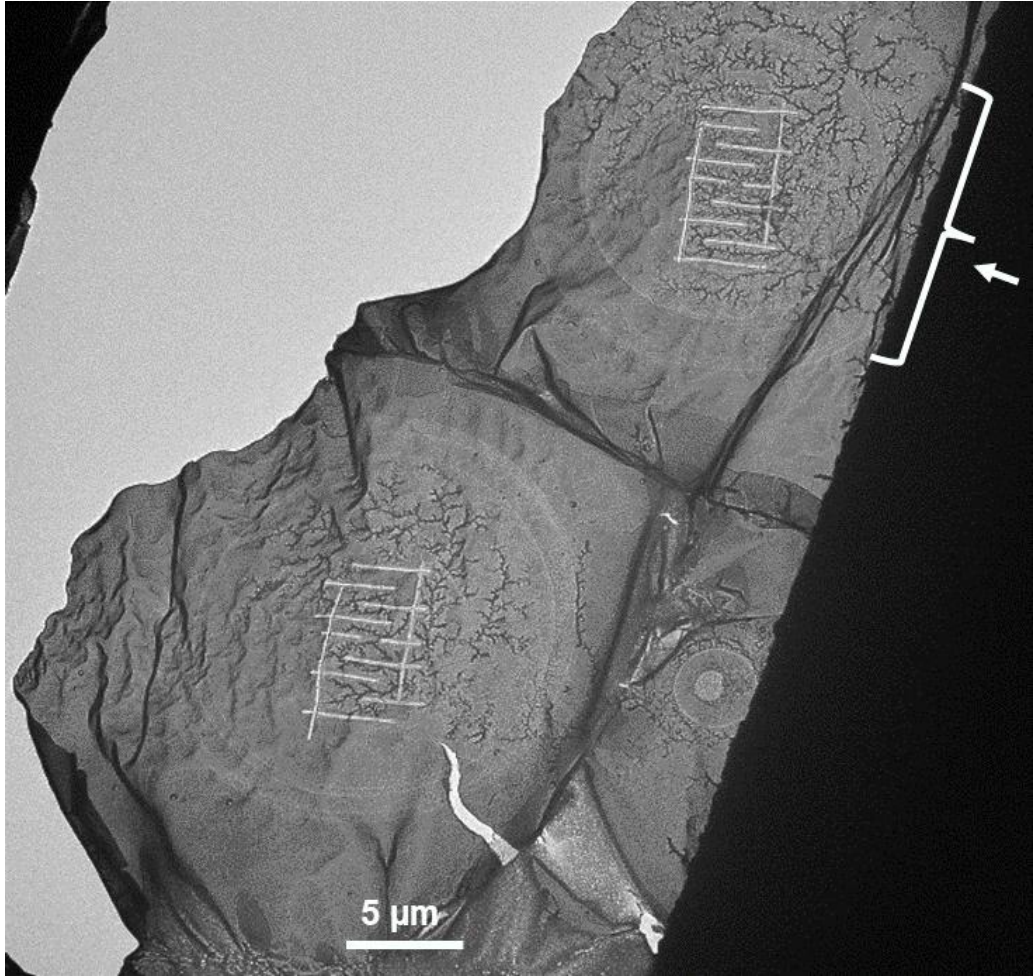

Figure 8. A TEM image of the patterned BPD/Ag nanosheet showing the stemming area of fractal formation at the edge near the Cu-grid.

While there is limited knowledge regarding charge propagation in semiconducting and insulating materials, previous studies have shown that electrons' pathways exhibit a fractal behaviour [35]. For instance, Richardella *et al*. [46] observed the distribution of electrons across the surface of manganese-doped gallium arsenide using a specialized scanning tunnelling microscope, which exhibited interconnected fractal-like patterns. Skinner *et al*. [47] confirmed in a theoretical study that the fractal ordering of low energy sites enhances the hopping conductivity of electrons, leading to their fractal propagation in semiconductors.

In this paper, we introduce the concept of "electron-leader" (EL) to describe the flow of negative electrons from the Cu-grid to the positively charged nanosheet due to bond cleavage and the release of $Ag^+$ ions upon irradiation. The formation of the metallic fractal pattern began with the propagation of EL, which spread in a branching tree-like pattern in search of the most conductive path in the nanosheet. The branches of the metallic fractal pattern began to form and grow as Ag$^+$ ions diffuse onto the oppositely charged fractal path created by the EL.

Unlike the well-known diffusion limited aggregation (DLA) model in which the formation and growth of fractal patterns is directed by the random diffusion of particles in the surrounding medium [32], the formation of fractal nanostructures in this study is guided by a dynamic diffusion process, in which the metallic ions nucleate and grow following the EL path over long distances. We coin this process as dynamic diffusion aggregation (DDA). Consequently, the growth rate of the cluster is determined by the rate at which the EL can propagate through the medium. This explains the gradual nucleation and growth of NPs instead of their sudden formation. Therefore, we propose that the formation of metallic fractal nanostructures in the BPD nanosheet was determined primarily by the negatively charged EL path and secondarily by the DDA of Ag$^+$ along the EL paths.

### *2.4 Patterning Through the Utilization of Electron-Beam Lithography (EBL)*

Using the results obtained thus far and the subsequent understanding of the e-beam/nanosheet interactions, EBL was employed to further prove the direct patternability of the printed nanosheets and the possibility of creating complex more sophisticated patterns. In this regard, another conjugated dithiol/Ag nanosheet was printed with BDMT as the molecular organic backbone. BDMT has a conjugated molecular backbone and therefore is expected to behave similarly to the BPD-based nanosheet. For ease of handling, the printed nanosheet was placed on a TEM Cu-grid before subjecting it to EBL in order to easily transfer the nanosheet and image it by TEM after lithography. The EBL was performed on the BDMT/Ag nanosheet using a low does of 20000 μC/cm$^2$. This dose is significantly lower compared to the e-beam doses used in the previous in-situ TEM experiments and hence the formation of metallic fractals under EBL at this energy level was not anticipated, details on the EBL process can be found in the Methods section. The TEM images in Fig. 9 show the high-resolution details of the nanostructures fabricated on BDMT/Ag nanosheet using EBL. Upon examination of the TEM maping image in Fig. 9b, several key features standout. The well-defined edges and interfaces of the created nanostructures suggest that the EBL process was carried out with high precision, ensuring minimal deviation from the intended design. Moreover, the nanosheet maintained its stability under e-beam irradiation, as expected from the previous results and discussion in section 2.2. This is a significant achievement considering the challenges associated with fabricating and handling sensitive molecular based materials at such a small scale. It is also important to note that once the EBL patterned nanosheet was exposed to the TEM's high energy e-beam irradiation, the formation of fractal nanostructures was observed, see red arrows in Fig. 9C.

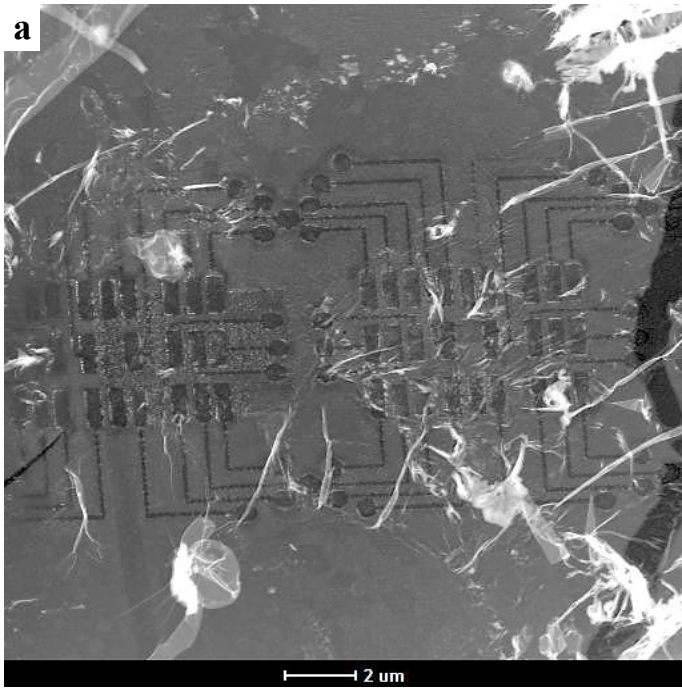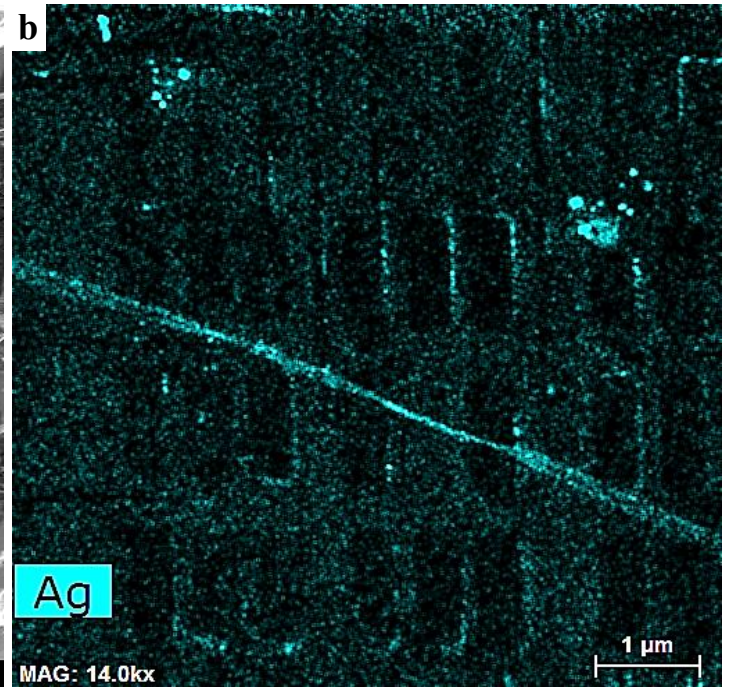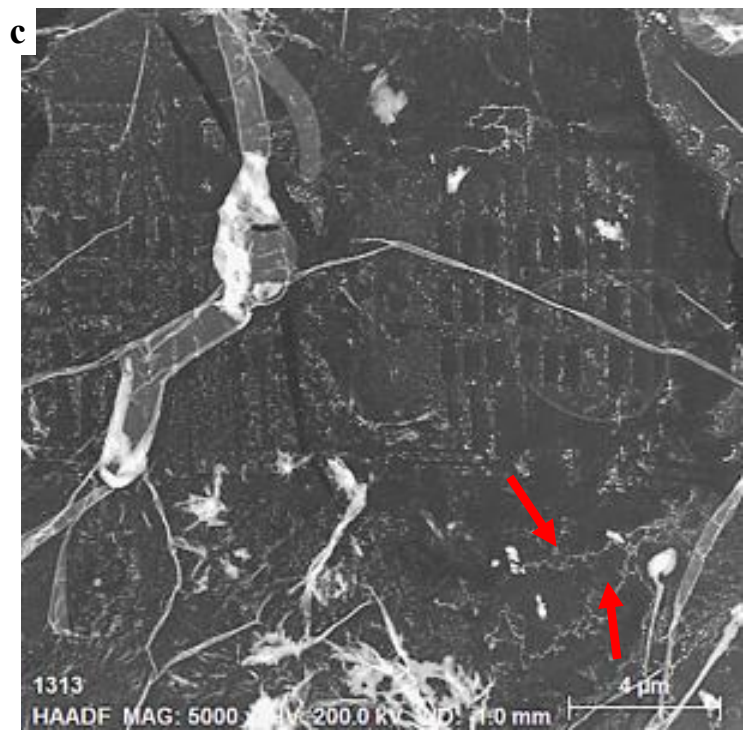

Figure 9. (a) DF-TEM image, (b) STEM-HAADF mapping of Ag in the BDMT/Ag nanosheet after EBL patterning, and (c) the formation of fractal nanostructures after TEM e-beam irradiation,

EBL process. Thus, introducing a novel direct-patterning lithography process that eliminates the need for the use of masks or polymer resists. The entire prenting process relies on the intrinsic interactions between the printed C/metal hybrid 2D-nanosheets and the e-beam. Furthermore, no follow-up metal deposition steps are required since the metal precourser is embedded within the nanosheet during the initial printing process, which later during e-beam irradiation, forms unique nano-fractal nanostructures (electrodes) directly on the patterned nanosheet. This resist and metal deposition-free process simplifies fabrication and ensures the material remains uncontaminated, which is vital for sensitive electronic applications. This method offers a cost-effective and highly precise approach for creating intricate structures on ultra-thin nanosheets.

Moving forward, our aim is to integrate the fractal metallic nanostructures as antennas within the molecular diodes, which are the C-based molecular building blocks of the nanosheet. Molecular diodes are critical for energy conversion and electronic applications, and their integration with the fractal electrodes could lead to the development of highly efficient nanoelectronic systems. This integration would allow for the conversion of energy at the molecular level, potentially leading to breakthroughs in energy-harvesting technologies and next-generation electronic devices.

## 3. Conclusion

In the last few decades, e-irradiation has been introduced as a new method for the fabrication of nanostructures and has been extensively used to prepare nano-scale clusters and materials. In this work, we report the direct synthesis of metallic nano-fractal networks directly on hybrid Technomolecular carbon/Ag nanosheets under high-energy e-beam irradiation. We used in-situ TEM irradiation experiments and compared the use of saturated C9 and conjugated BPD molecules as the backbone of the hybrid Technomolecular nanosheets to elucidate the impact of molecular electronic properties on the direct nucleation, growth, and formation of Ag fractal nanostructures. Our results demonstrated that the conjugated BPD/Ag nanosheet formed complex fractal nano-networks, which could be precisely controlled to achieve different patterns. Whilst the saturated C9/Ag nanosheet exhibited the formation of localized metallic nanoparticles. The different behaviour is investigated and attributed to the different molecular and electronic properties of both nanosheets. We further demonstrate the feasibility of resist-free electron-beam lithography (EBL) to fabricate complex circuit patterns directly on BDMT/Ag nanosheets, eliminating traditional resists and metal deposition. This approach enables precise, uncontaminated designs for flexible electronics, bioelectronics, and energy harvesting, including fractal antennas for rectennas, wearable bio-sensors, and unclonable security identifiers. By integrating molecular systems with fractal electrodes, Technomolecular materials could revolutionize nanotechnology, from sustainable energy to next-generation computing.

## 4. Methods

### 4.1 Materials and 3D Printing of Nanosheets

Two metalorganic carbon/metal nanosheets were synthesized using saturated 1,9-nonane-dithiol (C9) and conjugated (5,5'-bis(mercaptomethyl)-2,2'-bipyridine) or (BPD) as building-block units to examine the effect of the molecular structure of the carbon-based matrix on the formation of the metallic nanostructures under electron beam irradiation. Both saturated and conjugated-based nanosheets of (C9/Ag) and (BPD/Ag), respectively, were synthesized using the molecular self-assembly 3D printing technique using C9 or BPD molecules as the building-block units and Ag atoms as the mediator connecting the dithiol monolayers together. Details on the materials used and the 3D printing synthesis process can be found in ref [13]. After printing, the metalorganic nanosheets were fished from the printing solution using a silicon substrate and washed with DI water to remove any excess $Ag^+$ salts. For direct patterning by EBL, a third nanosheet of conjugated 1,4-Benzenedimethanethiol (BDMT) moleucles was printed and fished using the exact same molecular 3D-printing method.

*4.2 Characterization*

To investigate the structural order of the as-printed nanosheets, Fourier-transform infrared spectroscopy (FTIR) was performed using a ThermoScientific (Nicolet iS50 FT-IR) spectrometer. The as-printed nanosheets were deposited after cleaning on KBr pellets to conduct FTIR measurements in transmittance mode. FTIR spectra were recorded with 128 scans per sample/background in the spectral range of 4000–400cm$^{-1}$ at 4cm$^{-1}$ spectral resolution.

To assess the current density of both the as-printed saturated and conjugated nanosheets and explore how the molecular structure influences the formation of fractal nanostructures, we employed a scanning probe microscope (SPM). The obtained results were averaged across ten successful junctions. The procedure involved depositing the as-printed nanosheets onto an Si substrate, followed by vacuum drying. Our SPM equipment is sourced from NT-MDT Spectrum Instruments. In current spectroscopy, we measure the current passing through the probe, known as the "Iprobe signal," as a function of the voltage applied between the probe and the sample. It's important to note that current measurements are exclusively applicable to conductive samples. We conducted these measurements using a SCM-PIT-V2 probe with a conductive platinum coating from Bruker. The I(BV) plot furnishes insights into the local conductivity of the sample under investigation.

*4.3 In-situ Electron-Beam Irradiation and Nanoparticle Formation*

In this study, the e-beam within the TEM was used for both imaging and inducing metallic nanostructures formation, thus allowing for *in-situ* observations during synthesis. A CETA M-camera (Thermo Fisher Scientific Inc, USA) and Axon studio software were used was used to capture and record an image every second during the analysis, and the collective images were combined together to make the movies in the supporting information, thus facilitating the *in-situ* electron irradiation experiment. The electron dose (current density) was varied during the analysis according to the patterning protocol being studied. In general, two irradiating protocols were applied. In protocol 1, the electron dose started low at ~800 A/m$^2$ and then increased gradually up to ~150,000A/m$^2$ to induce metallic nanostructures formation and patterning in the irradiated areas. In protocol 2, the nanosheets were exposed to a sudden high dose of 80,000-150,000 A/m$^2$ in order to expel the Ag atoms and induce its precipitation and patterning outside the illuminated areas. Protocol 2 was used to create and carve different paths and patterns on the 2D nanosheets, followed by protocol 1 to induce metallic nanostructured formation. Eventually, the growth location/path of the metallic nanostructures was controlled based on the created patterns. It is also important to note that the intensity can vary briefly at the initial exposure to the electron beam during the time period required to focus for imaging (a few seconds). Both stationary and moving e-beams were utilized in the patterning of the nanosheets.

*4.4 Design of Nano-trap Pathways Using In-situ Electron-Beam Irradiation*

An example of a complex nano-trap pattern created using e-beam can be seen in Fig. 10. The trap consisted of straight lines making a rectangle and several dot-like features inside it. Initially, a high-dose electron beam (~100,000A/m$^2$) was focused on point 1 and moved along the yellow arrows as indicated in Fig. 10 to make the rectangle. However, the last line was not fully closed or attached to the starting point, see point 2 in Fig. 10, in an aim to maintain an entrance for the metallic fractals and trap its growth into the rectangle area.

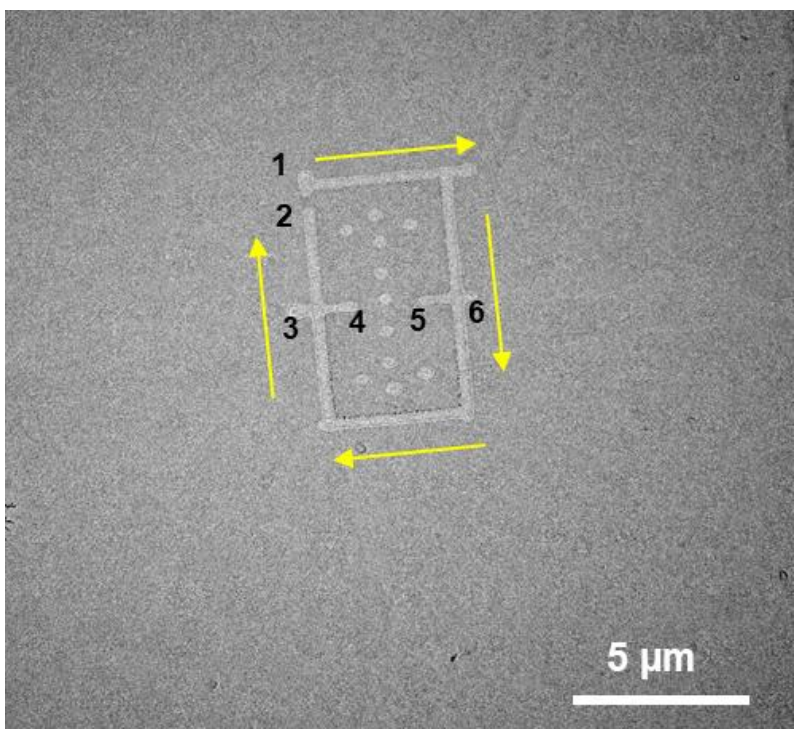

Figure 10. BF-TEM image showing the sequence of steps with which the trap-like complex pattern was created using e-beam on the BPD/Ag nanosheet.

After that, the electron beam was blanked to prevent any unwanted irradiation induced interactions with the nanosheet. In a similar process, the two shorter lines on the two sides of the rectangle were fabricated by focusing the blanked beam on point 3, un-blanking it and moving it to point 4, followed by blanking the beam and moving it to point 5, un-blanking it, and moving it to point 6. Next, the electron beam was moved inside the rectangle to the desired coordinates and the dot-like features were fabricated simply by focusing

S21

and un-blanking the e-beam on the desired spot on the nanosheet. The process was repeated until the desired pattern was created.

*4.5 Direct Patterning by Electron-Beam Lithography of BDMT/Ag Nanosheet*

EBL was employed to further prove the direct patternability of the printed nanosheets and the possibility of creating complex more sophisticated patterns. For this test, a conjugated BDMT/Ag nanosheet was printed using the molecular 3D-printing method and then deposited directly on a 3 mm TEM Cu-grid. This is to facilitate the handling of the nanosheet during patterning under EBL and its transfer to the TEM after EBL for further examination. An Elphy Multibeam attachement from Raith compagny on the FEI Versa 3D system was used for this experiment. The patterning parameters are a beam current of 0.031 nA, area dose of 20000 $\mu C/cm^2$ and an area dwell time of 0.645 ms. The Writefield size was 100 μm with a step size of 2.5nm.

**Supporting Information Available:**

Figures S1-S7: Additional microscopic, spectroscopic, measurements and data analysis conducted on both 3D-printed conjugated and saturated nanosheets (PDF).

Movies V1-V8: Recorded videos of the 3D printing process and the *in-situ* TEM irradiation of both saturated and conjugated nanosheets (MP4).

**Author Contributions**

H.H designed the experiments, H.H and S.A prepared the nanosheets, H.H and A.Z conducted TEM, STM, and SEM experiments. Data analysis is conducted by H.H, S.A, V.A.E, Y.T, and K.Y.T. All authors contributed to manuscript preparation and reviewing processes. H.H Proposed the Technomolecular materials and DDA, H.H directed the project.

**Acknowledgement**

H.H discloses support for the research of this work from HBKU Thematic Research Grant Program [VPR-TG01-005]. Authors acknowledge the support of the Core Labs at Hamad bin Khalifa University. Authors Acknowledge Dr. Kamal Mroue from the Core Labs at Hamad bin Khalifa University for conducting the FTIR experiment.

**Competing Interests**

The authors declare no competing financial interests.

**Data Availability**

The data that supports the findings of this study are available from the corresponding author upon reasonable request, email: hhamoudi@hbku.edu.qa

# References

[1] Hiesinger, P.R. The Self-Assembling Brain: How Neural Networks Grow Smarter, Princeton University Press, 2021

[2] Dehaene, S. Seeing the Mind: Spectacular Images from Neuroscience, and What They Reveal about Our Neuronal Selves. The MIT Press, Basic Books, 2023.

[3] Afshinmanesh, F.; A.G. Curto, A.G.; Milaninia, K.M.; Van Hulst, N.F.; and Brongersma, M.L. Transparent metallic fractal electrodes for semiconductor devices. Nano letters. 2014, 14 (9) 5068-5074. DOI: 10.1021/nl501738b

[4] Li, G.; Chen, X.; Ni, B.; Li, O.; Huang, L.; Jiang, Y.; Hu, W.; and Lu, W. Fractal H-shaped plasmonic nanocavity, Nanotechnology. 2013, 24 (20) 205702. DOI: 10.1088/0957-4484/24/20/205702.

[5] Gottheim, S.; Zhang, H.; Govorov, A.O.; and Halas, N.J. Fractal Nanoparticle Plasmonics: The Cayley Tree, ACS Nano. 2015, 9 (3) 3284. DOI: 10.1021/acsnano.5b00412

[6] Yun, J.; Lee, H.; Song, C.; Jeong, Y.R.; Park, J.W.; Lee, J.H.; Kim, D.S.; Keum, K.; Kim, M.S.; Jin, S.W.; Lee, Y.H.; Kim, J.W.; Zi, G.; and Ha, J.S. A Fractal-designed stretchable and transparent micro supercapacitor as a Skin-attachable energy storage device, Chemical Engineering Journal. 2020, 387 (1) 124076. DOI: 10.1016/j.cej.2020.124076

[7] Watterson, W.J.; Montgomery, R.D.; and Taylor, R.P. Fractal Electrodes as a Generic Interface for Stimulating Neurons, Scientific Reports. 2017, 7 (1) 6717. DOI: 10.1038/s41598-017-06762-3

[8] Karmakar, A. Fractal antennas and arrays: a review and recent developments, International Journal of Microwave and Wireless Technologies. 2021, 13 (2) 173. DOI: 10.1017/S1759078720000963

[9] Roe, E.T.; Bies, A.J.; Montgomery, R.D.; Watterson, W.J.; Parris, B.; Boydston, C.R.; Sereno, M.E.; and Taylor, R.P. Fractal solar panels: Optimizing aesthetic and electrical performances, PloS one. 2020, 15 (3) e0229945. DOI: 10.1371/journal.pone.0229945

[10] Xiang, D.; Wang, X.; Jia, C.; Lee, T.; and Guo, X. Molecular-Scale Electronics: From Concept to Function, Chemical Reviews. 2016, 116 (7) 4318. DOI: 10.1021/acs.chemrev.5b00680

[11] Hamoudi. H.; and Esaulov, V.A. Selfassembly of α,ω-dithiols on surfaces and metal dithiol heterostructures, Annalen der Physik. 2016, 528 (1) 242. DOI: 10.1002/andp.201500280

[12] Ariga, K. Nanoarchitectonics: The method for everything in materials science, Bulletin of the Chemical Society of Japan. 2024, 97 (1) 1-26. DOI: 10.1093/bulcsj/uoad001

[13] Hamoudi, H.; Berdiyorov, G.R.; Zekri, A.; Tong, Y.; Mansour, S; Esaulov, V.A.; and Youcef-Toumi, K. Building block 3D printing based on molecular self-assembly monolayer with self-healing properties, Scientific Reports. 2022, 12 (1) 6806. DOI: 10.1038/s41598-022-10875-9

[14] Sola, A.; Trinchi, A.; and Hill, A.J. Self-assembly meets additive manufacturing: Bridging the gap between nanoscale arrangement of matter and macroscale fabrication, Smart Materials in Manufacturing. 2023, 1 (1) 100013. DOI: 10.1016/j.smmf.2022.100013

[15] Hamoudi, H.; Guo, Z.; Prato, M.; Dablemont, C.; Zheng, W.Q.; Bourguignon, B.; Canepa, M.; and Esaulov, V.A. On the self-assembly of short-chain alkanedithiols, Physical Chemistry Chemical Physics. 2008, 10 (1) 6836. DOI: 10.1039/B809760G

[16] Aped, I.; Mazuz, Y.; and Sukenik, C.N. Variations in the structure and reactivity of thioester functionalized self-assembled monolayers and their use for controlled surface modification, Beilstein J Nanotechnol. 2012, 3 (1) 213. DOI: 10.3762/bjnano.3.24

[17] Kohli, P.; Taylor, K.K.; Harris, J.J; and Blanchard, G.J. Assembly of Covalently-Coupled Disulfide Multilayers on Gold, Journal of the American Chemical Society. 1998, 120 (46) 11962. DOI: 10.1021/ja981987w

[18] Ouellette, R.J.; and Rawn, J.D. Organic Chemistry: Structure, Mechanism, Synthesis *(Second Edition)*, Academic Press, 2019, DOI: 10.1016/C2016-0-04004-4.

[19] Zhang, Q.; Peng, X.; Nie, Y.; Zheng, Q.; Shangguan, J.; Zhu, C.; Bustillo, K.C.; Ercius, P.; Wang, L.; Limmer, D.T.; and Zheng, H. Defect-mediated ripening of core-shell nanostructures, Nature Communications. 2022, 13 (1) 2211. DOI: 10.1038/s41467-022-29847-8

[20] Longo, E.; Avansi, W.; Bettini, J.; Andrés, J.; and Gracia, L. In situ transmission electron microscopy observation of Ag nanocrystal evolution by surfactant free electron-driven synthesis, Scientific reports. 2016, 6 (1) 1. DOI: 10.1038/srep21498

[21] Gong, J.; Liu, H.; Jiang, Y.; Yang, S.; Liao, X.; Liu, Z.; and Ringer, S. In-situ synthesis of Ag nanoparticles by electron beam irradiation, Materials Characterization. 2015, 110 (1) 1. DOI: 10.1016/j.matchar.2015.09.030

[22] Andres, J.; Longo, E.; Gouveia, A.; Costa, J.P.C.: Gracia, L.; and Oliveira, M. In situ formation of metal nanoparticles through electron beam irradiation: modelling real materials from first-principles calculations, Journal of Material Sciences & Engineering. 2018, 7 (3) 1000461. DOI: DOI: 10.4172/2169-0022.1000461

[23] Chen, Q.; Dwyer, C.; Sheng, G.; Zhu, C.; Li, X.; Zheng, C.; and Zhu, Y. Imaging Beam-Sensitive Materials by Electron Microscopy, Advanced Materials. 2020, 32 (16) 1907619. DOI: 10.1002/adma.201907619

[24] Hamoudi, H.; Chesneau, F.; Patze, C.; and Zharnikov, M. Chain-Length-Dependent Branching of Irradiation-Induced Processes in Alkanethiolate Self-Assembled Monolayers, The Journal of Physical Chemistry C. 2011, 115 (2) 534. DOI:10.1021/jp109434k

[25] LaMer, V.K.; and Dinegar, R.H. Theory, Production and Mechanism of Formation of Monodispersed Hydrosols, Journal of the American Chemical Society. 1950, 72 (11) 4847. DOI: 10.1021/ja01167a001

[26] Kamachali, R.D. Melting upon Coalescence of Solid Nanoparticles, Solids. 2022, 3 (2) 361. DOI: 10.3390/solids3020025

[27] Turchanin, A.; and Gölzhäuser, A. Carbon nanomembranes from self-assembled monolayers: Functional surfaces without bulk, Progress in Surface Science. 2012, 87 (5-8) 108. DOI: 10.1016/j.progsurf.2012.05.001

[28] Ballav, N.; Schilp, S.; and Zharnikov, M. Electron-Beam Chemical Lithography with Aliphatic Self-Assembled Monolayers, Angewandte Chemie. 2008, 120 (8) 1443. DOI: 10.1002/anie.200704105

[29] Khan, I.; Huang, S.; and Wu, C. Multi-walled carbon nanotube structural instability with/without metal nanoparticles under electron beam irradiation, New Journal of Physics. 2017, 19 (1) 123016. DOI: 10.1088/1367-2630/aa969c

[30] Azcárate, J.C.; Fonticelli, M.H. and Zelaya, E. Radiation damage mechanisms of monolayer-protected nanoparticles via TEM analysis, The Journal of Physical Chemistry C. 2017, 121 (46) 26108. DOI: 10.1021/acs.jpcc.7b08525

[31] Schindelin, J.; Arganda-Carreras, I.; Frise, E.; Kaynig, V.; Longair, M.; Pietzsch, T.; Preibisch, S.; Rueden, C.; Saalfeld, S.; Schmid, B.; Tinevez, J.Y.; White, D.J.; Hartenstein, V.; Eliceiri, K.; Tomancak, P.; and Cardona, A. Fiji: an open-source platform for biological-image analysis, Nature Methods. 2012, 9 (1) 676. DOI: 10.1038/nmeth.2019

[32] Witten, T.A.; and Sander, L.M. Diffusion-limited aggregation, Physical Review B. 1983, 27 (9) 5686. DOI: 10.1103/PhysRevB.27.5686

[33] Taylor, R. Vision of beauty, Physics World. 2011, 24 (5) 22. DOI: 10.1088/2058-7058/24/05/33

[34] Tong, Y.; Alsalama, M.; Berdiyorov, G.R.; and Hamoudi, H. A combined experimental and computational study of the effect of electron irradiation on the transport properties of aromatic and aliphatic molecular self-assemblies, Nanoscale Adv., 2022, 4 (1) 3745-3755. DOI: 10.1039/D2NA00040G

[35] Ding, Z.; Li, C.; Da, B.; and Liu, J. Charging effect induced by electron beam irradiation: a review, Science and Technology of Advanced Materials. 2021, 22 (1) 932. DOI: 10.1080/14686996.2021.1976597

[36] Cazaux, J.; Correlations between ionization radiation damage and charging effects in transmission electron microscopy, Ultramicroscopy. 1995, 60 (3) 411. DOI: 10.1016/0304-3991(95)00077-1

[37] Chen, Y.T.; Wang, C.Y.; Hong, Y.J.; Kang, Y.T.; Lai, S.E.; Chang, P.; and Yew, T.R. Electron beam manipulation of gold nanoparticles external to the beam, RSC Advances. 2014, 4 (1) 31652. DOI: 10.1039/C4RA03350G

[38] Sabri, M.M.; and Möbus, G. New insight into nanoparticle precipitation by electron beams in borosilicate glasses, Applied Physics A. 2017, 123 (1) 455. DOI: 10.1007/s00339-017-1064-5

[39] Sidorov, A.I.; Nashchekin, A.V.; Nevedomskiy, V.N.; Usov, O.A.; and Podsvirov, O.A. self-assembling of silver nanoparticles in glasses under electron beam irradiation, International Journal of Nanoscience. 2011, 10 (6) 1265. DOI: 10.1142/S0219581X11008411

[40] Jiang, N. Beam damage by the induced electric field in transmission electron microscopy, Micron. 2016, 83 (1) 79. DOI: 10.1016/j.micron.2016.02.007

[41] Jiang, N. Damage by Induced Electric Field in Beam-sensitive Materials, Microscopy and Microanalysis. 2017, 23 (S1) 1812. DOI:10.1017/S1431927617009722

[42] Hauwiller, M.R.; Zhang, X.; Liang, W.I.; Chiu, C.H.; Zhang, Q.; Zheng, W.; Ophus, C.; Chan, E.M.; Czarnik, C.; and Pan, M. Dynamics of nanoscale dendrite formation in solution growth revealed through in situ liquid cell electron microscopy, Nano letters. 2018, 18 (10) 6427. DOI: 10.1021/acs.nanolett.8b02819

[43] Tang, J.; Li, Z.; Xia, Q.; and Williams, R.S. Fractal structure formation from Ag nanoparticle films on insulating substrates, Langmuir. 2009, 25 (13) 7222. DOI: 10.1021/la9010532

[44] Cherington, M.; McDonough, G.; Olson, S.; Russon, R.; and Yarnell, P.R. Lichtenberg figures and lightning: case reports and review of the literature, Cutis. 2007, 80 (2) 141. PMID: 17944174.

[45] Feynman, R.P.; Leighton R.B.; and Sands, M. The Feynman Lectures on Physics, The New Millennium Edition: Mainly Electromagnetism and Matter, Volume 2, Basic Books, 2011.

[46] Richardella, A.; Roushan, P.; Mack, S.; Zhou, B.; Huse, D.; Awschalom, D.; and Yazdani, A. Visualizing Critical Correlations near the Metal-Insulator Transition in Ga1-xMnxAs, Science. 2010, 327 (5966) 665. DOI: 10.1126/science.1183640

[47] Chen T.; and Skinner, B. Enhancement of hopping conductivity by spontaneous fractal ordering of low-energy sites, Physical Review B. 2016, 94 (8) 085146. DOI: 10.1103/PhysRevB.94.085146

# Supporting Information for

**Technomolecular Materials: 3D-Printed 2D-Nanosheets with Self-Patterned Electrodes**

Hicham Hamoudi*, Sara Iyad Ahmad, Atef Zekri, Kamal Youcef-Toumi, Vladimir A. Esaulov

**Affiliations:**

H. Hamoudi*, S. I. Ahmad, A. Zekri: Qatar Environment and Energy Research Institute, Hamad Bin Khalifa University, 34110, Doha, Qatar.

*Corresponding author Email: hhamoudi@hbku.edu.qa, hichamhamoudia@gmail.com

K. Youcef-Toumi: Mechatronics Research Laboratory, Massachusetts Institute of Technology, MA 02139, USA.

Vladimir A. Esaulov: Institut des Sciences Moléculaires d'Orsay, UMR 8214 CNRS-Université, bât 520, Université Paris Sud, Université Paris Saclay, Orsay 91405, France.

**This PDF file includes:**

Supplementary Figures 1 to 7

Captions for Supplementary Figures 1 to 7

Captions for Movies V1 to V8

**Other Supplementary Materials for this manuscript include the following:**

Movies V1 to V8

**Supplementary Figures**

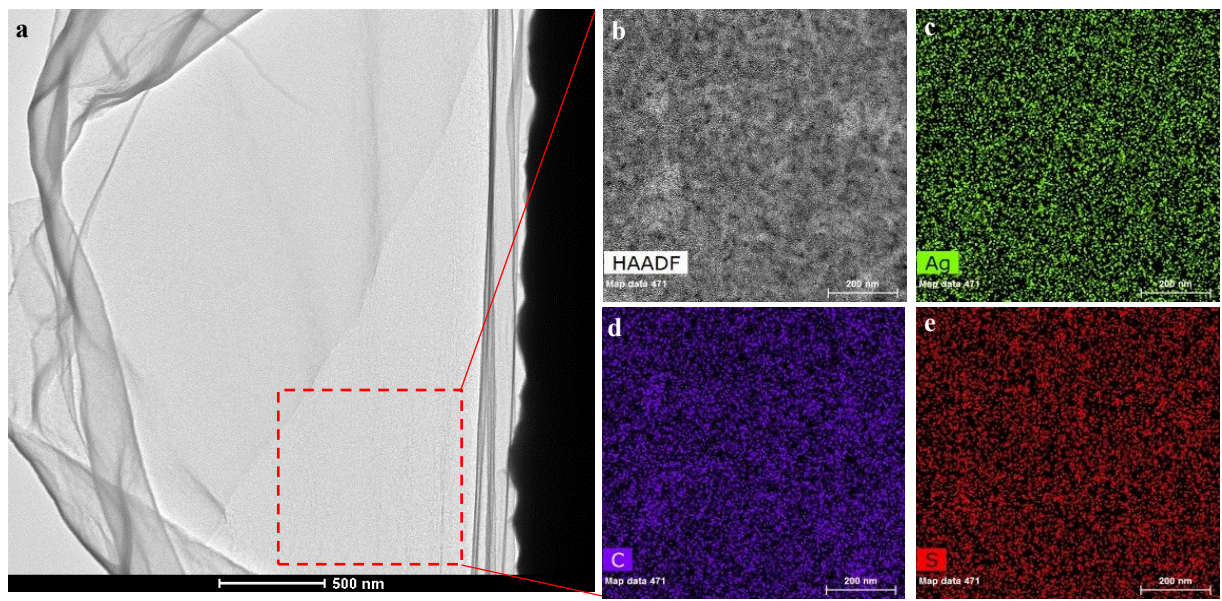

*Figure S1.* a) BF-TEM image of the as-printed hybrid C9-dithiol/Ag nanosheet, and (b-e) HAADF elemental mapping of Ag, C, and S, respectively.

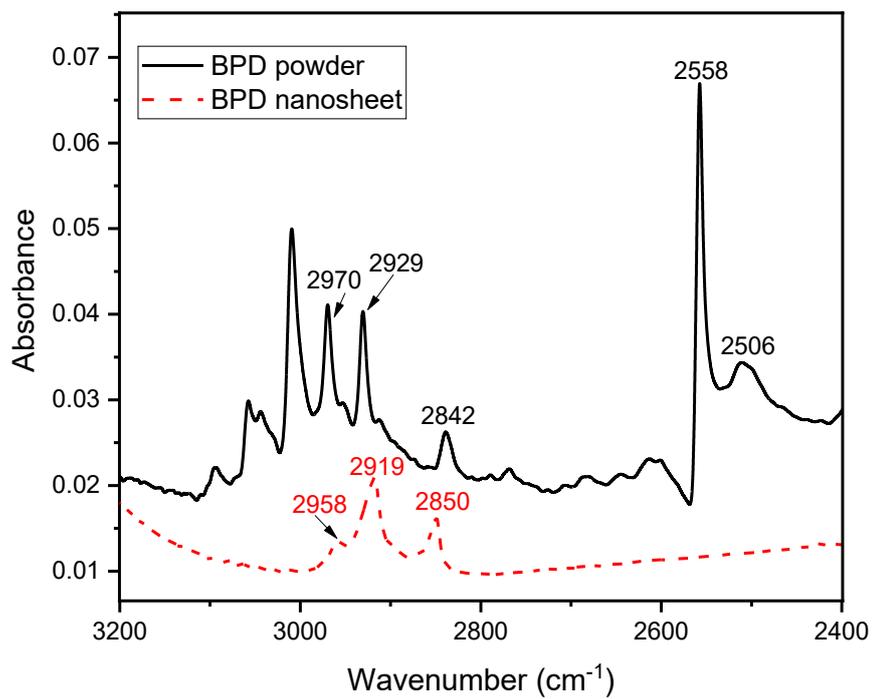

*Figure S2.* FTIR spectra of BPD reference in powder (solid black line) and as-printed BPD freestanding nanosheet (dashed-red line) both on KBr pellets.

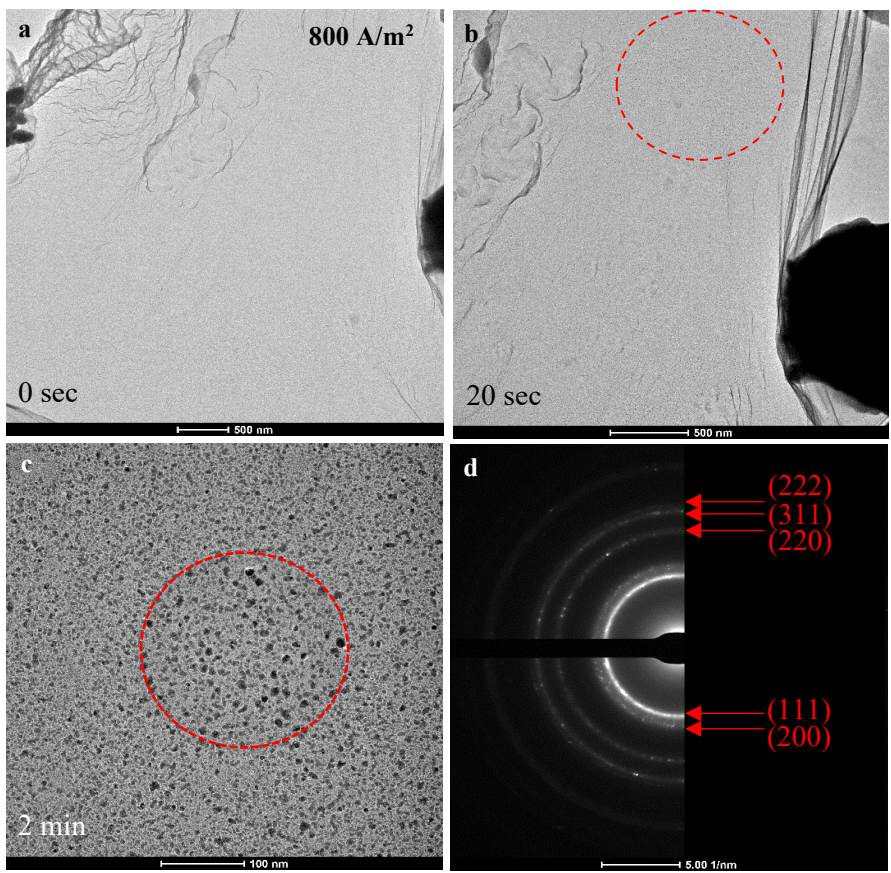

*Figure S3.* BF-TEM images of the printed sheet taken (a) immediately after e-beam exposure and (b) after exposure with an e-dose of 800A/m$^2$ for 20 seconds in the red-circled areas. (c) A magnified image of the area in red-circle 1 in (b) after the red-circle in (c) was exposed to e-dose of 1000A/m$^2$ for 2 minutes, and (d) is the SAED pattern of red-circle in (c).

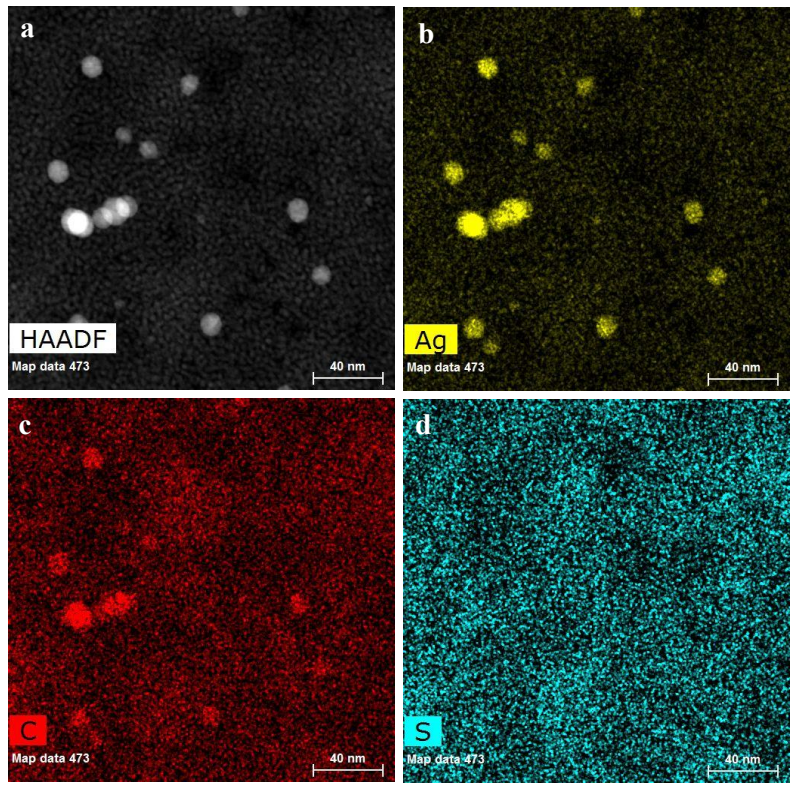

*Figure S4. (a) STEM-HAADF image from the irradiated areas in the C9/Ag nanosheet and its corresponding elemental mapping of (b) Ag, (c) C, and (d) S.*

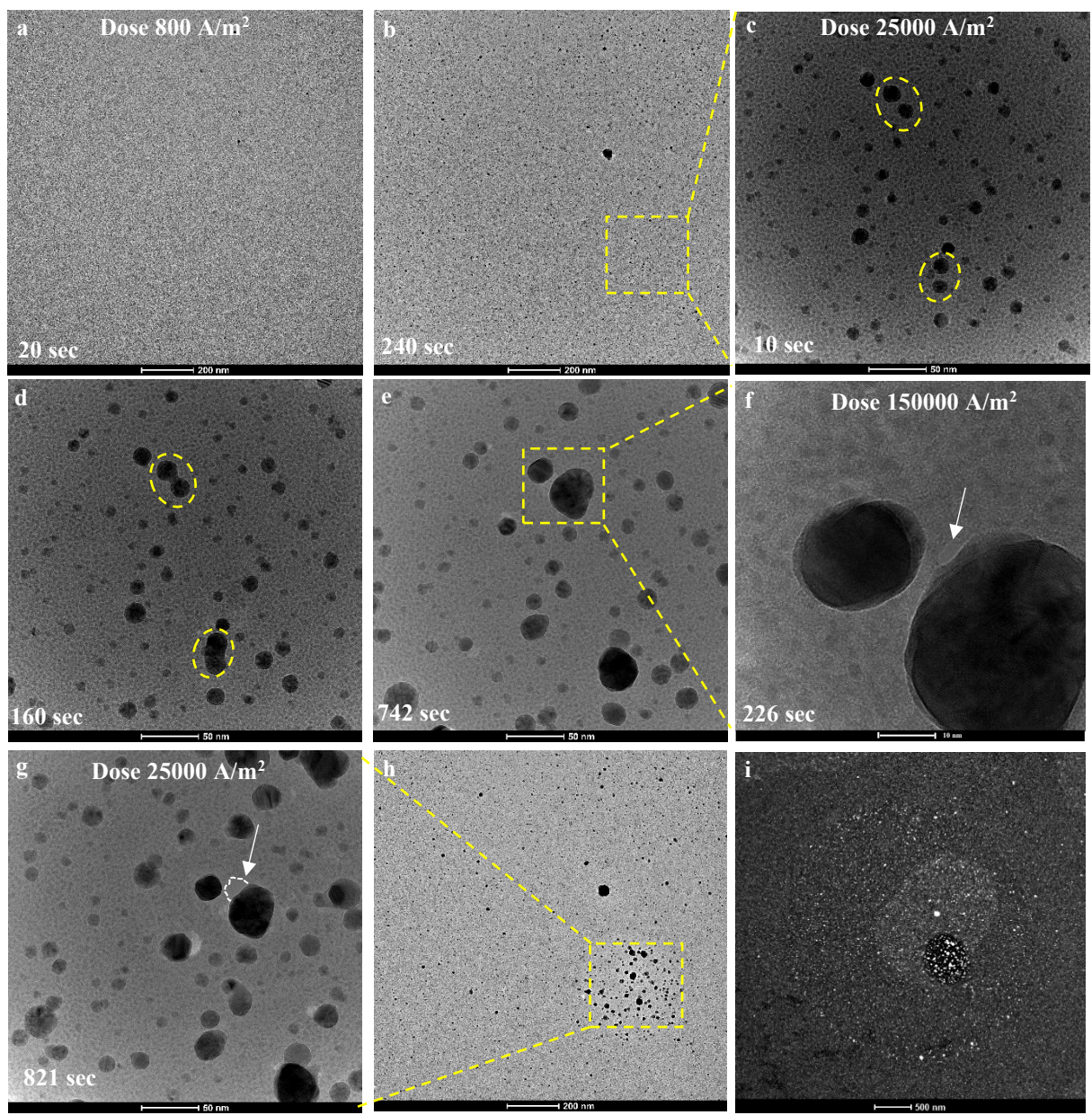

*Figure S5.* A sequence of TEM images that correspond to irradiating the C9-dithiol/Ag nanosheet with electron-beam at different magnifications of the same area. Irradiation time and magnification related electron-dose are recorded on each image. Irradiation dose in (a-b) 800 A/m$^2$, (c-e) 25000 A/m$^2$, (f) 150000 A/m$^2$, and (g) 25000 A/m$^2$. (c) Magnification of box in (b), (f) magnification of box in (e), (g) magnification of box in (h), and (i) dark-field image of the initially irradiated area in (a).

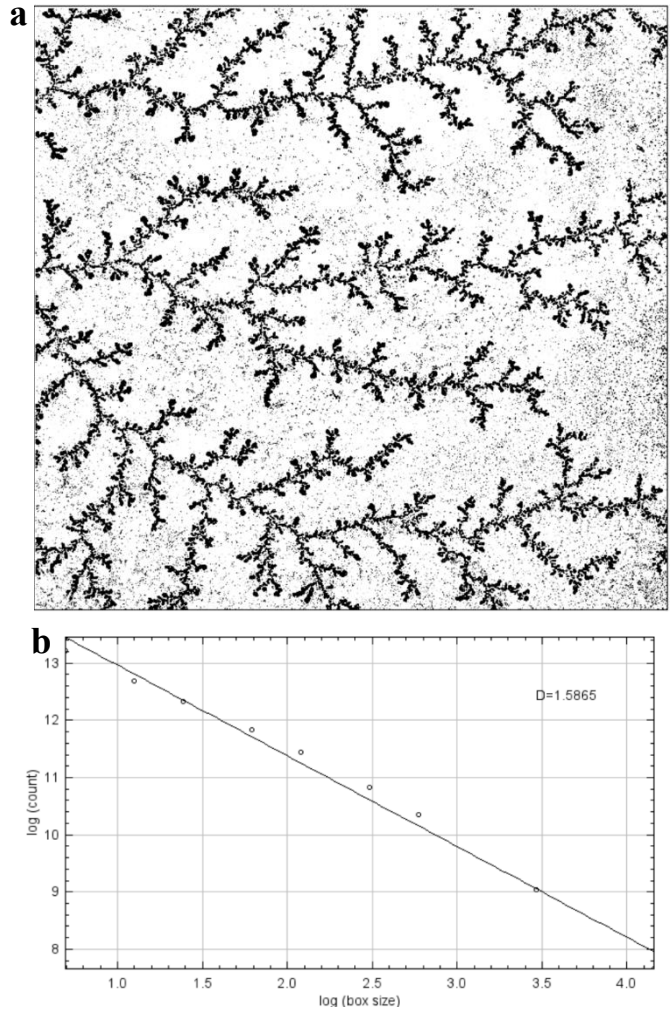

*Figure S6. The fractal dimension of the metallic fractal pattern shown in Fig. (a) is ~1.6 as determined by the (b) box counting method using the Fiji-ImageJ software.*

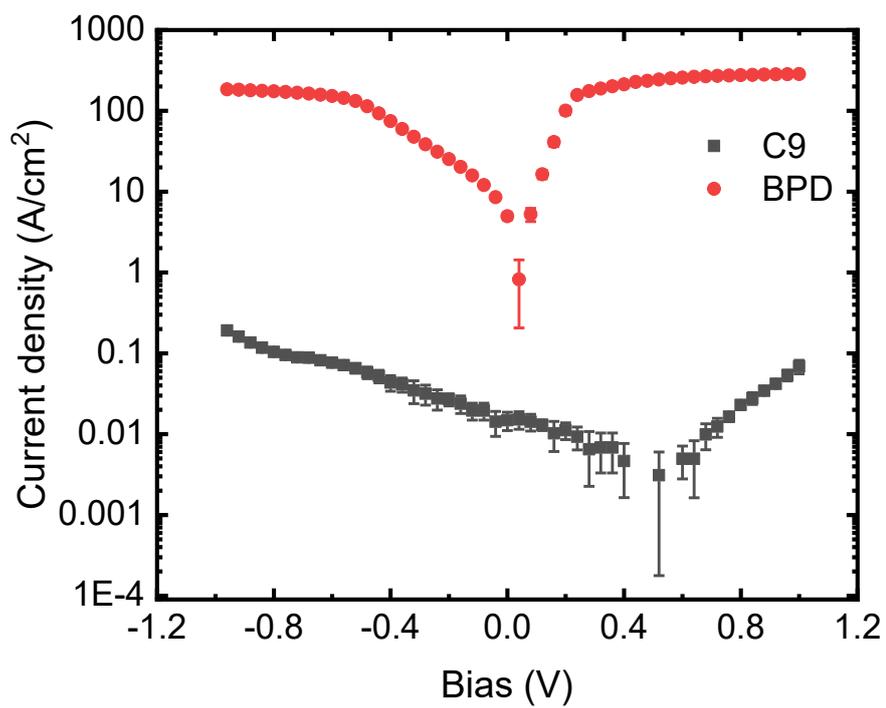

*Figure S7. Current-voltage characteristics of both BPD/Ag nanosheet (red-circle curve) and C9/Ag nanosheet (black-square curve).*

**Captions for Supplementary Movies**

**Movie V1.** 3D Printing of the Saturated C9/Ag Nanosheet, Fig. 1(a-c)

**Movie V2.** In-situ TEM Irradiation of C9/Ag Nanosheet, Fig. 5S(a-b)

**Movie V3.** In-situ TEM Irradiation of C9/Ag Nanosheet, Fig. 5S(c-e)

**Movie V4.** In-situ TEM Irradiation of C9/Ag Nanosheet, Fig. 5S(f)

**Movie V5.** In-situ TEM Irradiation of C9/Ag Nanosheet, Fig. 2(a-b)

**Movie V6.** In-situ TEM Irradiation of BPD/Ag Nanosheet Showing Circular Metallic Fractal Nano-electrodes, Fig. 5a

**Movie V7.** In-situ TEM Irradiation of BPD/Ag Nanosheet Showing Linear Metallic Fractal Nano-electrodes, Fig. 5b

**Movie V8.** In-situ TEM Irradiation of BPD/Ag Nanosheet Showing Complex Metallic Fractal Nano-electrodes, Fig. 5c